\documentclass[a4paper]{article}

\usepackage{INTERSPEECH2020}
\usepackage{amsmath, amsfonts,color}
\usepackage{xcolor}

\usepackage{hyperref}
\hypersetup{
    colorlinks=true,
    linkcolor=blue,
    filecolor=magenta,
    urlcolor=blue,
    citecolor=blue,
}

\title{Controllable neural text-to-speech synthesis using intuitive prosodic features}

\name{Tuomo~Raitio, Ramya~Rasipuram, Dan Castellani}

\address{Apple}

\email{}

\begin{document}

\maketitle

\begin{abstract}
Modern neural text-to-speech (TTS) synthesis can generate speech that is indistinguishable from natural speech. However, the prosody of generated utterances often represents the average prosodic style of the database instead of having wide prosodic variation. Moreover, the generated prosody is solely defined by the input text, which does not allow for different styles for the same sentence. In this work, we train a sequence-to-sequence neural network conditioned on acoustic speech features to learn a latent prosody space with intuitive and meaningful dimensions. Experiments show that a model conditioned on sentence-wise pitch, pitch range, phone duration, energy, and spectral tilt can effectively control each prosodic dimension and generate a wide variety of speaking styles, while maintaining similar mean opinion score ($4.23$) to our Tacotron baseline ($4.26$).
\end{abstract}


\section{Introduction}

State-of-the-art neural text-to-speech (TTS) technology commonly consists of a sequence-to-sequence (seq2seq) neural network with attention \cite{NIPS2014_5346, wang2017tacotron, shen2017natural} that maps text or phone input to a Mel-spectrogram, and a neural back-end \cite{oord2016wavenet, kalchbrenner2018efficient, kumar2019melgan} that converts the Mel-spectrogram into a sequence of speech samples. These two networks are trained with a large amount of speech data, enabling the generation of speech that can be indistinguishable from natural speech \cite{shen2017natural}.

The prosody of the generated speech is largely defined by the seq2seq encoder--decoder neural network, in TTS commonly referred to as Tacotron \cite{wang2017tacotron, shen2017natural}. However, the prosody of the generated utterances often represents the average style of the database instead of having a wide range of prosodic variation. Moreover, the generated prosody is defined by the input text, which does not allow for different speaking styles for the same sentence. Therefore, the generated prosody may not always correspond to the intended message for the listener. To generate a wide range of prosodic output that is appropriate for the context, the Tacotron model must learn to factorize the various prosodic styles in the training data, and also provide a meaningful control over the prosodic output independent of the input text.



Traditionally, prosody modeling is based on schematizing and labeling prosodic phenomena and developing rule-based systems or statistical models from the derived data \cite{Prosody2010}. However, the prosodic attributes are difficult and time-consuming to annotate, and therefore such an approach is impractical for modern TTS with large databases. In contrast, end-to-end neural TTS systems learn prosody as an inherent part of the model, which makes the generation of speech with the correct linguistic content and the appropriate prosody a consistent modeling task. However, such prosodically unsupervised models lack the explicit control over the output prosody.

Recent developments in end-to-end neural TTS have provided some solutions to prosody modeling. In \cite{skerryryan2018endtoend}, an extension to the Tacotron architecture is introduced to capture the residual attributes that are not specified by the linguistic input. The model learns a latent embedding space of prosody through conditional auto-encoding, derived from the target Mel-spectrograms, to enable prosody transfer.


As an improvement to \cite{skerryryan2018endtoend}, \cite{Wang2018StyleTU} proposed global style tokens (GSTs) to learn a clustered latent space of prosody. The same prosody encoder as in \cite{skerryryan2018endtoend} is used, but a style token layer is added where the prosody embedding is used as a query to an attention layer that maps the embedding to the GSTs using combination weights. The output of the style token layer is a style embedding that describes the prosodic style. The architecture can be used for either prosody transfer using a reference audio or for directly generating a specific prosodic style by choosing a style token.


The aforementioned unsupervised methods for learning a latent space of prosody have many advantages. Unsupervised methods do not need any extra information or labeling of prosody, and all the components are jointly trained to achieve a consistent model. However, there are some drawbacks. First, the latent space represents all the residual acoustic differences in the reference Mel-spectrograms after linguistic content has been accounted for. As a result, any acoustic differences due to channel (recording setup, noise type and level) and (undesirable) variability in speaker's voice from one day to another are also learned. Therefore, instead of focusing on meaningful prosodic aspects of speech, the model learns an entangled representation of prosody and unknown underlying acoustic factors. Second, manual listening of the tokens is required to determine what sort of style tokens the model has learned from the data, and how to utilize these for reproducing a desired prosody. Third, the number of style tokens must be set heuristically, and the model is not guaranteed to find perceptually distinct and prosodically meaningful clusters for each of the tokens. The individual style tokens are often either similar to each other, or it may be hard to make practical use of the learned style tokens as they may not represent any desired styles.




The prosody of language covers all aspects of speech that are not related directly to the articulation for the linguistic expression. Although there is no agreed number of prosodic variables, there is a set of variables that is widely agreed to have a major contribution to prosody. These variables are 1) pitch, 2) length of the speech sounds, 3) loudness, and 4) timbre or voice quality \cite{Campbell2003vq,Sluijter1996SpectralBA}. These four subjective auditory variables can be accurately measured using the following acoustic variables: fundamental frequency, phone duration, speech energy, and spectral tilt, respectively. Assuming that the four acoustic features cover the prosodic space, then there are certain benefits in using them for prosody modeling and control: 1) they are disentangled so that they can be independently varied, 2) they are intuitive so that it is easy to understand their effect on prosody, 3) they are independent from other acoustic factors, such as background noise or other recordings conditions, thus making them robust for prosody modeling.

In this work, we take a similar, unified end-to-end approach for prosody modeling as in \cite{skerryryan2018endtoend, Wang2018StyleTU}, but instead of performing conditional auto-encoding using the Mel-spectrograms, we use prosodically meaningful acoustic features derived from the speech signal, similar to \cite{Shechtman_2019}. This approach has the benefit of guiding the model to learn only perceptually relevant acoustic differences that contribute to prosody. We predict the acoustic features directly from the encoder outputs, similar to \cite{stanton2018predicting}, but with a more focused set of prosodic features, which enables direct and intuitive control over each prosodic dimension.

Our main contribution in this work is a unified and simple neural network architecture for prosody modeling and control using intuitive prosodic features. In contrast to \cite{skerryryan2018endtoend, Wang2018StyleTU, stanton2018predicting}, we use well-defined acoustic features for prosody modeling instead of Mel-spectrograms. Also, in contrast to \cite{Shechtman_2019}, we use a simple unified encoder-decoder architecture that is trained as a single model. While prosodic features such as fundamental frequency, phone duration, and speech energy, have seen increased interest for prosody modeling in neural TTS \cite{Shechtman_2019,wan2019chive,klimkov2019finegrained,sun2020fully}, we extend our model to utilize the fourth dimension of prosody: spectral tilt \cite{Campbell2003vq,Sluijter1996SpectralBA}. We show that each feature has a specific and effective control over the prosodic space, and that the proposed method can generate various speaking styles while maintaining high quality\footnote{Speech samples can be found at \href{https://apple.github.io/neural-tts-with-prosody-control/}{\nolinkurl{https://apple.github.io/neural-tts-with-prosody-control/}}.}.

\section{Technical Overview}

The baseline sequence-to-sequence with attention model is similar to the Tacotron 2 model \cite{shen2017natural}. The input is a phoneme sequence with punctuation and word boundaries, and the output is a Mel-spectrogram. Our experiments show that using location-sensitive monotonic attention \cite{raffel2017online} or location-sensitive stepwise monotonic attention \cite{he2019robust} results in a more robust alignments than global alignment. For consistency, the former is used throughout this study. We use similar mechanism to \cite{shen2017natural} for utterance end-point prediction. We also use a streaming architecture to reduce the lag in the output. We are able to fit three Tacotron models in a single GPU, each of them generating Mel-spectrograms 5x real-time, while our on-device implementation runs 8.5x real-time on a mobile device.

To generate a sequence of speech samples from the Mel-spectrogram, we use an autoregressive recurrent neural network (RNN) based architecture, similar to WaveRNN \cite{kalchbrenner2018efficient}. The model consists of a single RNN layer with 512 hidden units, conditioned on Mel-spectrogram, followed by two fully-connected layers ($512\times256$, $256\times256$), with single soft-max sampling at the output. The model is trained with pre-emphasized speech sampled at 24~kHz and $\mu$-law quantized to 8 bits for efficiency. We can run three of these models in parallel on a single GPU, each of them generating speech 7.7x real-time, while our on-device implementation runs 3.3x real-time on a mobile device.

More information about the architecture and the on-device implementation of the baseline system can be found in \cite{appleneuraltts2020}.



\subsection{Proposed Prosody Control Architecture}

\begin{figure}[t]
  \centering
  \includegraphics[width=1.0\linewidth]{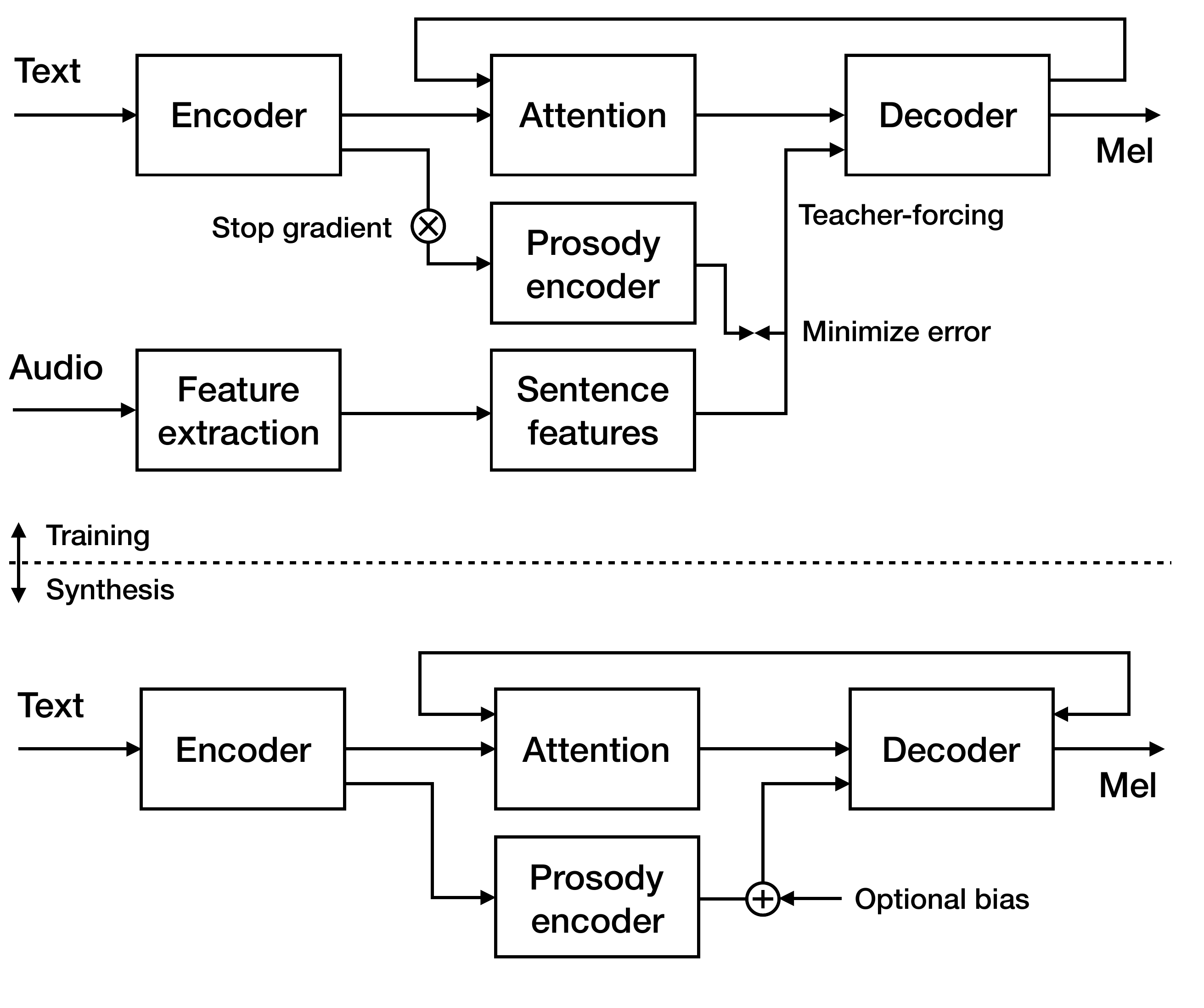}
  \vspace{-8mm}
  \caption{An overview of the proposed prosody modeling encoder-decoder with attention architecture. Top part describes the training phase where the prosody encoder learns to predict the sentence-wise prosodic features, and the decoder is conditioned on the ground-truth features (teacher-forcing). The bottom part describes the inference phase where prosody encoder predicts prosodic features to condition the decoder, with an additional bias option for prosody control.}
  \vspace{-4mm}
  \label{fig:architecture}
\end{figure}

We extend the baseline architecture by introducing a prosody encoder \cite{skerryryan2018endtoend} that predicts prosodic features from the encoder outputs. The prosody encoder consists of a stacked 3-layer LSTM with hidden state size of 128 at each layer. The last state of the LSTM is fed to a fully connected projection layer with a tanh non-linearity. The input to the prosody encoder is the phone embedding, and the output is the prosody feature vector that is concatenated with the decoder input for prosody control. To jointly train the model with both phone and prosody encoders, we add a stop gradient operation between the two modules. The stop gradient ensures that the training of the prosody encoder will not negatively affect the training of the phone encoder. We also use teacher-forcing of the prosodic features to efficiently train the network. The model architecture is illustrated in Fig.~\ref{fig:architecture}.

\subsection{Prosodic Features}
\label{sec:prosodic_features}

To learn a prosodically meaningful latent space for prosody control, we use acoustic features extracted from the original speech to condition the model. We use fundamental frequency (pitch), phone duration, speech energy, and spectral tilt to model the prosodic space. These features are easy to calculate from speech signals and are robust against background noise or other recordings conditions. These features are also disentangled to a large degree so that they can be independently varied. Overall, these features provide an intuitive control over the prosodic space.

We extract the pitch of speech using 3 pitch estimators, and then vote for the most likely pitch trajectory. We compute two features from the voted pitch contour: average log-pitch and log-pitch range of voiced speech. The latter is calculated as the difference between the 0.05 and 0.95 quantile values of the frame-wise log-pitch contour for each utterance. We use automatic speech recognition to force-align the text and audio to obtain the phone durations, and calculate the average log-phone duration per utterance. We extract the frame-wise log-energy\footnote{Speech energy $E$ is calculated as $E = 20log10(\hat{x})$, where $\hat{x}$ is the average absolute sample amplitude, excluding silence parts.} of each utterance (excluding silences), and calculate the frame-wise spectral tilt of voiced speech using first order all-pole modeling\footnote{Spectral tilt is measured using the predictor coefficient of the first order all-pole filter.}, and finally average them per utterance.


The utterance-wise acoustic features, log-pitch, log-pitch range, log-phone duration, log-energy, and spectral tilt, are then normalized to a range $[-1,1]$ by first calculating the median ($M$) and the standard deviation ($\sigma$) of each feature, and then projecting the data in the range [$M-3\sigma$, $M+3\sigma$] into $[-1,1]$. Finally, we clip values $|x|>1$ so that all data is in the range $[-1,1]$. This process is similar to the approach in \cite{Shechtman_2019}.

\section{Experiments}

\subsection{Data}

We use an internal 36-hour dataset of an American English female voice. In addition, we have a 16-hour conversational expressive dataset from the same speaker to expand the prosodic coverage. All recordings were produced in a professional studio and post-processed for high quality. The original speech data is down-sampled to 24~kHz for training the neural TTS system. 80-dimensional Mel-spectrograms are computed from pre-emphasized speech using short-time Fourier transform (STFT) with 25~ms frame length and 10~ms shift.

\subsection{Models}

We trained the following three models:

\vspace{-1mm}
\begin{itemize}
\setlength\itemsep{-0.5mm}
\item[1.] {\bf Baseline:} High-quality baseline model trained with the 36-hour dataset.
\item[2.] {\bf Prosody 36h:} Our proposed prosody control model trained with the 36-hour dataset.
\item[3.] {\bf Prosody 52h:} Our proposed prosody control model trained with the 36-hour dataset combined with the 16-hour conversational dataset for better prosodic coverage.
\end{itemize}
\vspace{-1mm}

For the prosody control models, we used pitch, pitch range, phone duration, speech energy, and spectral tilt as the conditioning features. We train all the models for 3 million steps using a single GPU and batch size of 16. All systems use the same back-end WaveRNN model \cite{appleneuraltts2020}, trained with the 36-hour dataset, to generate speech from the Mel-spectrograms.


\subsection{Objective Measures}
\label{sec:objective}

To measure how well the model can control each prosodic dimension, we synthesized speech at different points in the $[-1,1]$ scale. Each dimension was varied independently. As text material, we used 199 sentences with general text and responses typical to a voice assistant. A total of 16,517\footnote{16,517 $=$ 5 dimensions $\times$ 9 values $\times$ 199 samples $\times$ 2 systems $+$ 199 baseline samples $-$ (5$-$1) $\times$ 2 $\times$ 199 samples repeated at 0 bias.} utterances were synthesized. Then we compared how well the output speech reflects the given target prosodic bias by measuring the corresponding acoustic features from the synthetic utterances.

\begin{figure}[t]
  \centering
  \includegraphics[width=1.0\linewidth]{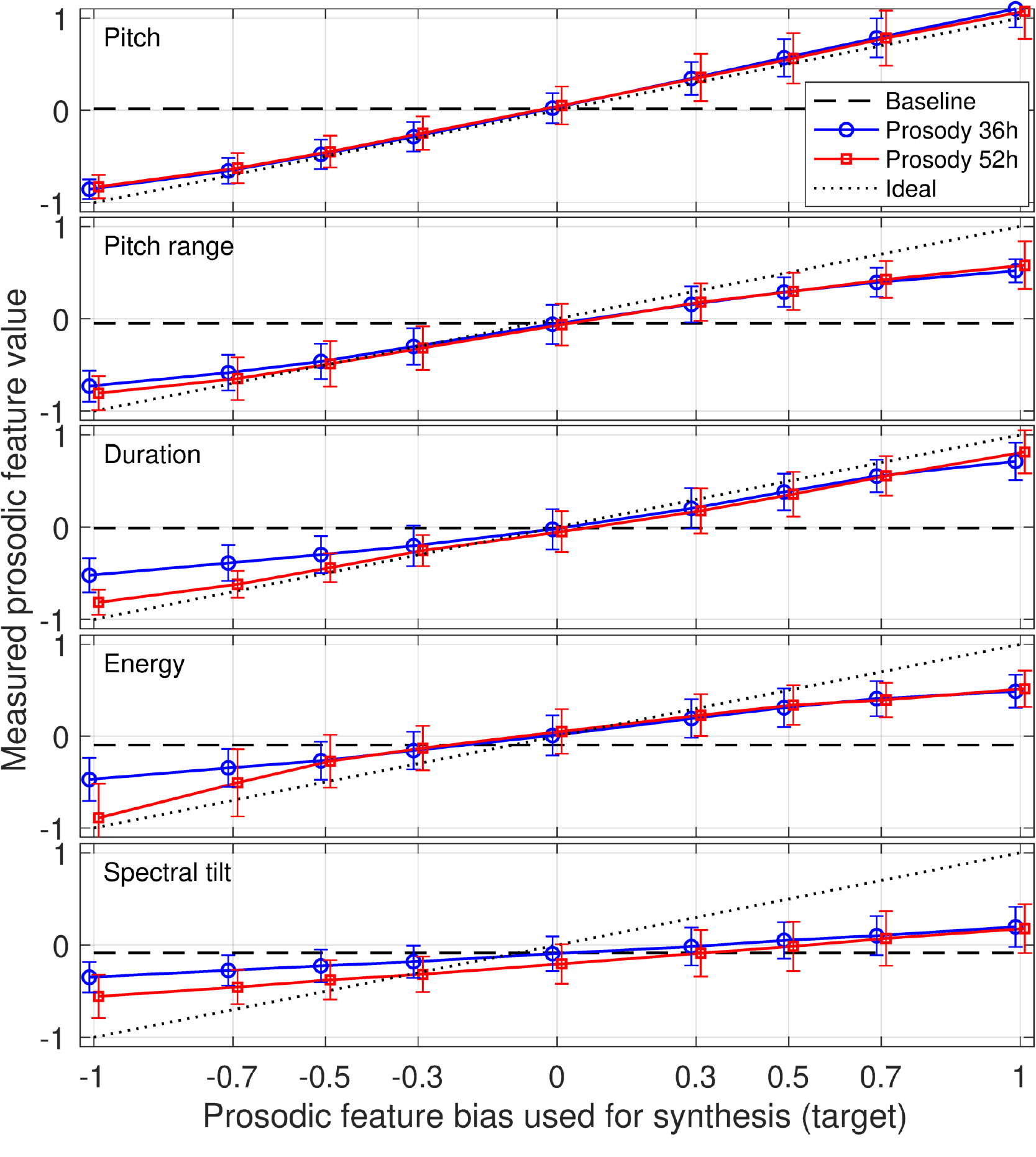}
  \vspace{-7mm}
  \caption{Means and standard deviations of the measured prosodic features with respect to target bias values.}
  \label{fig:measurements}
  \vspace{-2mm}
\end{figure}

Fig.~\ref{fig:measurements} shows the measured acoustic features at each target bias value, normalized by the procedure in Sec.~\ref{sec:prosodic_features}, using the original normalization values calculated over the whole database. The original acoustic feature values with respect to the scale $[-1,1]$ are shown in Table~\ref{tab:feature_values}. The results show that the target bias values of the prosodic features are well reflected in the output synthetic speech. Pitch shows almost ideal correlation between the target and measured values, whilst the remaining features show good correlation. The system trained with expressive speech (Prosody 52h) shows better correlation at the low-end of duration, energy, and spectral tilt.

\begin{table}[t]
  \caption{Prosodic feature values in normalized scale $[-1,1]$.}
  \footnotesize
  \vspace{-2mm}
  \label{tab:feature_values}
  \centering
  \begin{tabular}{|l|l|l|l|l|}
    \hline
    Feature & Unit & $-$1.0 & 0.0 & 1.0 \\
    \hline
    Pitch & Hz & 144.2 & 234.0 & 323.7 \\
    Pitch range & Hz & 50.9 & 355.8 & 660.8 \\
    Duration & ms & 32.7 & 117.6 & 202.5 \\
    Energy & dB & $-$26.2 & $-$20.7 & $-$15.2 \\
    Spectral tilt & - & $-$0.997 & $-$0.978 & $-$0.958 \\
    \hline
  \end{tabular}
  \vspace{-4mm}
\end{table}

\subsection{Listening Tests}

We carried out a number of listening tests to evaluate the subjective performance of the proposed approach. First, we evaluated the overall TTS naturalness of all the models without prosody control (zero bias). The intent was to assess the feasibility of the proposed architecture in place of the baseline system, even when manual prosody control is not specifically used.

From each system, we used the 199 synthetic utterances described in Sec.~\ref{sec:objective}. A 5-point mean opinion score (MOS) test was performed by 66 individual American English native speakers using headphones, resulting in a total of 8,986 responses. The results in Table~\ref{tab:mos_ab} show that all the three systems yield a high MOS, and that there is no statistically significant difference ($p > 0.10$) in quality between the three systems when listening to the speech samples in isolation. Therefore, we can conclude that the proposed prosody control systems can generate high quality synthetic speech.


\begin{table}[t]
  \caption{MOS (first row) and AB test results for the three systems without any manual prosody control. 95\% confidence intervals computed from the $t$-distribution are shown for MOS.}
  \footnotesize
  \label{tab:mos_ab}
  \vspace{-2mm}
  \centering
  \begin{tabular}{|l|l|l|l|}
    \hline
    \textbf{Baseline} & \textbf{Prosody 36h} & \textbf{Prosody 52h} & \textbf{No pref.} \\
    \hline
    4.26 ($\pm$ 0.030) & 4.23 ($\pm$ 0.032)    & 4.18 ($\pm$ 0.033) &  \\
    \hline
    48.3\%            & 25.9\%               & -                        & 25.9\%   \\
    43.5\%            & -                    & 25.7\%                   & 30.9\%   \\
    \hline
  \end{tabular}
  \vspace{-2mm}
\end{table}

MOS test can be somewhat insensitive to small differences in speech samples. For a more accurate assessment between the three systems, we performed AB listening tests where listeners were presented with a pair of speech samples, and they were asked to choose the one that sounded better, or choose no preference. We evaluated both prosody control models against the baseline. Ten proficient English speakers evaluated a total of 50 sample pairs using headphones, each test subject having a randomized set of different samples. The results, shown in Table~\ref{tab:mos_ab}, indicate that the baseline system is preferred over both prosody control models ($p < 0.005$). This degradation in quality could arise from the explicit prosody control that adds complexity, and thus makes it harder to make consistent prosody predictions. Also, matching the quality of our precisely tuned baseline system is a challenging task. The AB test also shows that the prosody control model trained with additional expressive speech is rated lower than the other models. This degradation could arise from the more challenging and varying speech material, which reduces the consistency of the output.

\begin{figure}[t]
  \centering
  \includegraphics[width=1.0\linewidth]{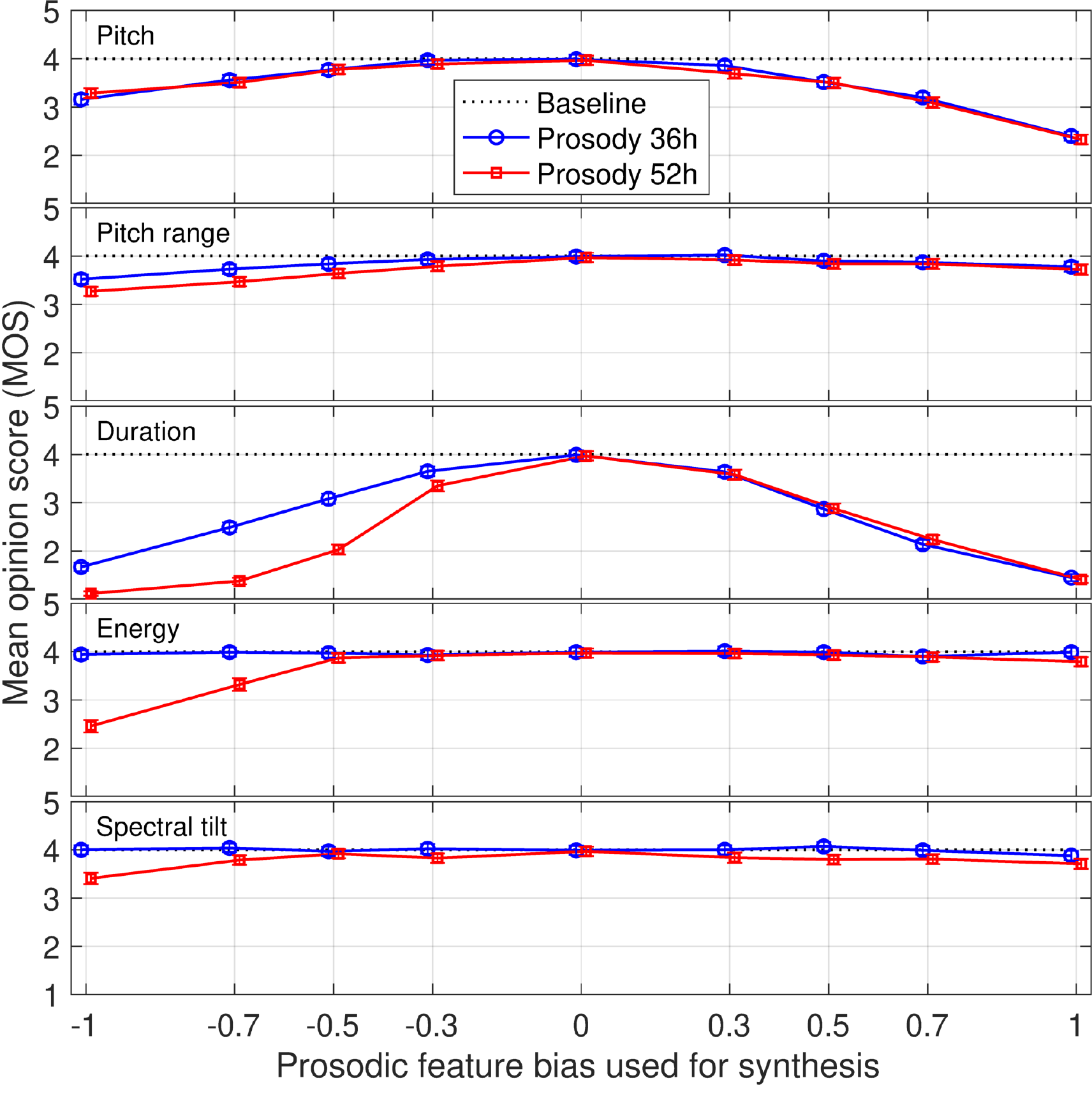}
  \vspace{-7mm}
  \caption{Means and 95\% confidence intervals of MOS measured over all features and target bias values.}
  \label{fig:mos}
  \vspace{-5mm}
\end{figure}

We also evaluated the MOS of the prosody control models at different points in the $[-1,1]$ feature scale. We used 40 sentences described in Sec.~\ref{sec:objective}. A total of 3,320 utterances were synthesized, and each was given 10 ratings, totaling 33,200 individual ratings by American English native speakers using headphones. The results in Fig.~\ref{fig:mos} show that the MOS has a tendency to slightly decrease towards both ends of the prosodic scale $[-1,1]$. This makes sense as the resulting speaking styles are less typical in comparison to the expected default delivery style of the voice talent. However, for some features, there is clear degradation in quality at the extreme ends of the scale as it results in speech production modes that do not occur in human speech. For example, when pitch bias is set to $1.0$, the utterance exhibits a sustained high pitch on the limits of the voice talent's voice. Similarly, very low or high phone duration in all the phones of an utterance does not correspond to normal speech production either. Due to wider prosodic range, the Prosody52h system shows more degradation at the extremes, especially with low energy where speech turns to almost whisper.

\subsection{Additional Experiments}



One of the aims of the prosody control architecture was to enable the generation of new improved, alternative, or more expressive versions of utterances that are key to the user experience of a voice assistant. We hypothesize that the intuitive control of prosodic features enables easy tuning of synthetic utterances. To test this hypothesis, we gathered 10 utterances that were lacking appropriate prosody for a voice assistant, consisting of either synthetic speech or original recordings from the studio. In a small proof-of-concept experiment, we presented dialog writers with a tool that had levers in the range $[-1,1]$ for each of the prosodic features, and let them generate new synthetic utterances by adjusting the prosodic dimensions. The Prosody 52h model was used for generating the samples in order to have greater prosodic flexibility. We then created a listening test to assess if the new speech samples were rated more appropriate than the existing ones. Listeners were presented with the original synthesis (or recording) and two new modified versions of the same utterance, and were asked to choose the one they preferred. Eight American English native speakers (different from those that created the data) performed the listening test. The prosody-modified samples were preferred 63\% of the time, and 44\% of the time even over original studio recordings. Although this test was very small in scale and without statistical power, it demonstrates a potential procedure of creating synthetic speech with improved prosody.

The best way to demonstrate the expressive capability of our proposed system is by listening. We present a set synthetic speech samples in \cite{raitio2020samples}, generated by varying the bias of each of the prosodic dimension gradually in the range $[-1,1]$.

\section{Discussion}

There are multiple benefits in the proposed architecture. First, the prosodic dimensions are inherently disentangled so that they can be independently varied. Second, the prosodic features are intuitive so that it is easy to understand their effect on prosody. Third, the prosodic features are independent from other acoustic factors such as background noise or other recordings conditions, thus making them robust for prosody modeling.

On the downside, the proposed model only provides sentence-level prosody control, and is lacking finer-grained control as in \cite{wan2019chive, sun2020fully}. Although the sentence-level features could be varied at the decoder input, accurate modification requires alignment information \cite{lee2019fine}. Alternatively, the prosodic features can be concatenated to the encoder output to enable phone-level prosodic modification. However, defining a good reference prosody becomes a problem, and the method would become similar to fine-grained prosody transfer \cite{klimkov2019finegrained}.

Like \cite{stanton2018predicting}, the proposed model predicts appropriate prosody directly from the text. Thus, the model can be used for synthesizing any text with appropriate style, provided with enough training data for the text and style. However, in our study, we did not achieve an overall better speech naturalness in comparison to our baseline system; sometimes the prosody was evaluated less natural in the AB test, indicating less stable prosody prediction for some text inputs and target style. For future work, we will investigate the robustness of the prosodic features, and explore different model architectures to overcome these issues.

\section{Conclusions}

We proposed a unified neural TTS system with prosody control capability using intuitive prosodic features. Subjective results show that the proposed model can synthesize speech with quality similar to our baseline, while being able to reproduce various prosodic styles. The objective results also show a clear correlation between the target prosodic feature values and the measured values from synthetic speech. Additional experiments demonstrate how the model can be used for creating new improved versions of synthetic speech with a desired prosody.


\newpage
\bibliographystyle{IEEEtran}

\bibliography{mybib}

\end{document}